\documentclass[preprint, nofootinbib, amsmath,amssymb,aps,]{revtex4-1}
\usepackage{graphicx}% Include figure files
\usepackage{dcolumn}% Align table columns on decimal point
\usepackage{bm}% bold math
\usepackage{color}
\usepackage[utf8]{inputenc}

\usepackage[normalem]{ulem}

\begin{document}
\preprint{APS/123-QED}
\title{Inequality, a scourge of the XXI century}
\author{José Roberto Iglesias}
\email{iglesias@if.ufrgs.br}
\address{Instituto de Física, Universidade Federal do Rio Grande
	do Sul, Porto Alegre, RS, Brazil}
\address{Escola de Gestão e Negócios, Programa de Pós-Graduação em
	Economia, UNISINOS, Porto Alegre, RS, Brazil}
\address{Instituto Nacional de Ciência e Tecnologia de Sistemas
	Complexos, INCT-SC, CBPF, Rio de Janeiro, RJ, Brazil}
\author{Ben-Hur Francisco Cardoso}
\email{ben-hur.cardoso@ufrgs.br}
\address{Instituto de Física, Universidade Federal do Rio Grande
	do Sul, Porto Alegre, RS, Brazil}
\author{Sebastián Gonçalves}
\email{sgonc@if.ufrgs.br}
\address{Instituto de Física, Universidade Federal do Rio Grande
	do Sul, Porto Alegre, RS, Brazil}
\address{URPP Social Networks, University of Zürich, Andreasstrasse 15, CH-8050 Zürich, Switzerland}

\date{\today}

\begin{abstract}

Social and economic inequality is a plague of the XXI Century.
It is continuously widening, as the wealth of a relatively small group increases and, therefore, the rest of the world shares a shrinking fraction of resources.
As an example, in 2016, the wealthiest 1\%  of US citizens possessed 40\% 
of the nation's wealth, while in 2007 they had less than 35\%~\cite{wolff2017household} ---considering the world's population, the richest 1\% have today 50\% of the total wealth.
This situation has been predicted and denounced by economists and econophysicists. 
The latter ones have widely used models of market dynamics which consider that wealth 
distribution is the result of  wealth exchanges among economic agents. 
A simple analogy relates the wealth in a society with the kinetic energy of 
the molecules in a gas, and the trade between agents to the energy exchange between
the molecules during collisions.
However, while in physical systems, thanks to the equipartition of energy, the gas eventually arrives at an equilibrium state, in many exchange models the economic system never equilibrates. Instead, it moves toward a ``condensed'' state, where one or a few agents concentrate all the wealth of the society and the rest of agents shares zero or a very small fraction of the total wealth.
According to Piketty~\cite{piketty2014capital}, many American countries, but also some European ones, are following this path.  Here we discuss two ways of avoiding the ``condensed" state. On one hand, we consider a  regulatory policy that favors the poorest agent in the exchanges, thus increasing the probability that the wealth goes from the richest to the poorest agent.
On the other hand, we study a tax system and its effects on wealth distribution. We compare the redistribution processes and conclude that complete control of the inequalities can be attained with simple regulations or interventions.
\end{abstract}

\maketitle

\section{Introduction}\label{intro}
Empirical studies of the distribution of income of workers, companies and countries were first presented, more than a century ago, by Italian economist Vilfredo Pareto. He asserted that in different European countries and times the distribution of income follows a power law behavior, i.e.  the cumulative distribution $P(w)$ of agents whose income is at least $w$ is given by $P(w) \propto w^{-\alpha}$~\cite{pareto1897cours}. Non-Gaussian distributions are denominated Levy distributions~\cite{stanley2000introduction}, thus this power law distribution is nowadays known as Pareto-Levy Distribution and the exponent $\alpha > 1$ is named Pareto exponent. The value of this exponent changes with location and time, but typical values are close to $3/2$. 

A power-law distribution implies a higher probability of having very rich agents compared with an exponential one ---also, higher probability of having very poor agents. According to Pareto, this means that inequality is not the result of chance, but different skills among agents. However, in 2003, Moshe Levy proved that power-law distributions are obtained even when all agents have almost the same winning probability; if one assumes different capabilities of increasing their wealth, the distribution obtained are not power law~\cite{levy2003investment}.

\begin{figure}
	\centering
	\includegraphics[width=0.7\textwidth]{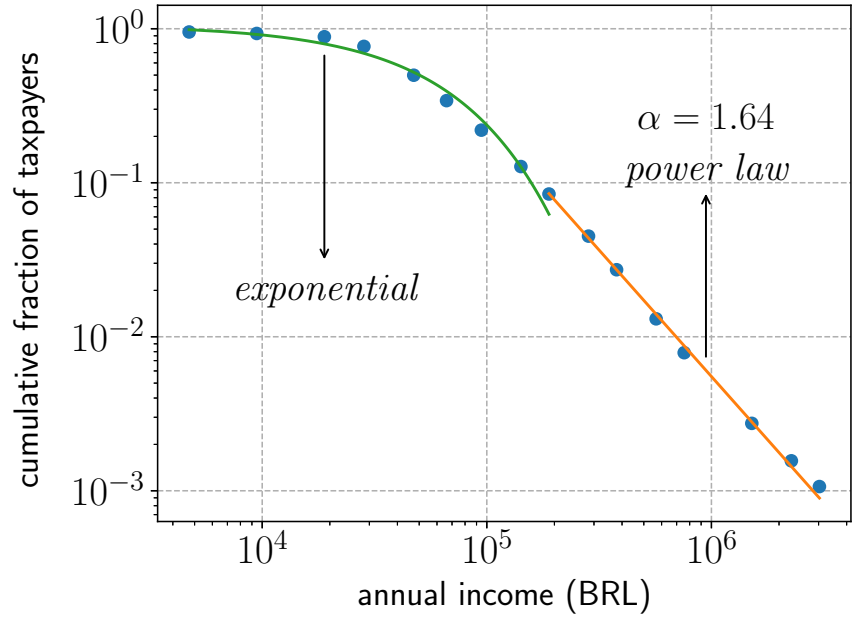}
	\caption{Cumulative fraction of taxpayers as a function of annual fiscal income, in Brazilian Real (BRL). Data collected from IRPF 2015~\cite{IRPF}.} 
	\label{fig:bra1}
\end{figure}

\begin{figure}
	\centering
	\includegraphics[width=0.7\textwidth]{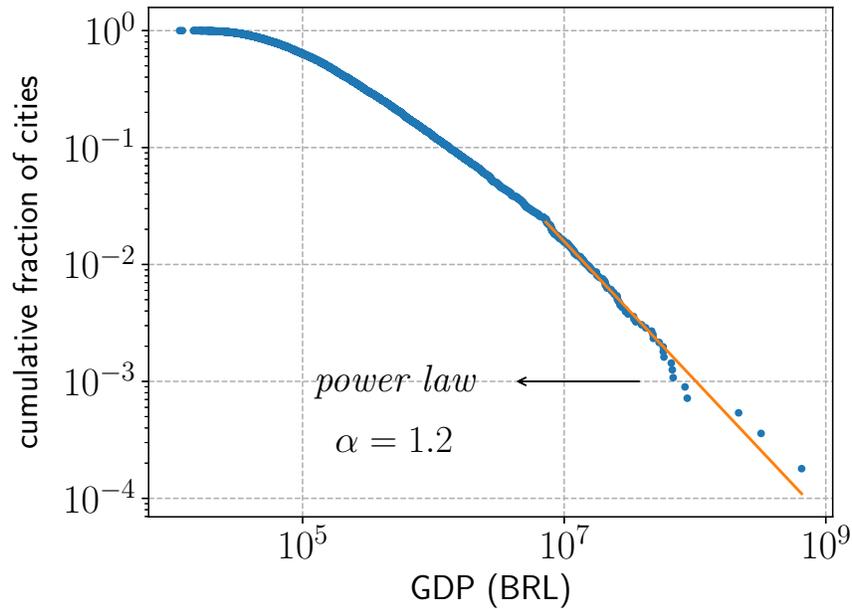}
	\caption{Cumulative fraction of cities as a function of annual fiscal Gros Domestic Product, in Brazilian Real (BRL). Data collected from IBGE 2015~\cite{IBGE}.} 
	\label{fig:bra2}
\end{figure}

Recent data indicates that, even though Pareto distribution provides a good fit to the distribution of high range of wealth, it fails in the middle and low range of wealth.  For instance, wealth data from Sweden and France~\cite{richmond2006review}, India~\cite{sinha2006evidence}, USA~\cite{klass2007forbes}, and UK~\cite{druagulescu2001exponential} can be fitted by a log-normal or exponential distribution for the low and middle range wealth, plus a power law for high wealth range. Similar behavior is also observed in the fiscal accumulated income distribution in Brazil, as shown in Fig.~\ref{fig:bra1}, where the Pareto exponent in the high-income region is equal to 1.64.
Or in the cumulative fraction of Brazilian cities, in terms of the accumulated annual Gross Domestic Product, shown in Fig.~\ref{fig:bra2}, with a tail Pareto exponent of 1.2.

A qualitative explanation for the two regimes is that Pareto considered data from income tax declarations, that did not apply to wages in the XIXth century. By considering high-income tax payers, he detected only wealth that grows in a multiplicative way, producing power-law tails.

In recent years, different agents-based or microscopic models of capital exchange among economic agents have been proposed to explain the empirical data (see ref.~\cite{caon2007unfair} for example). Most of these models consider an ensemble of interacting economic agents that exchange a fixed or random amount of their total ``wealth''. 
 The exchanged wealth is susceptible to several interpretations. It could be the money given for some service or commodity  or an ``error'' during the exchange~\cite{li2019affine}, or it may be attributed to a profit or ``plus valia''. If the exchanged amount is a  random fraction of the average wealth of the two agents, the resulting wealth distribution is ---unsurprisingly--- a Gibbs exponential distribution~\cite{yakovenko2009colloquium}.

Several variants of the basic model have been proposed to study  the distribution of wealth of different countries, generally under total wealth conservation. A well known proposition is that each agent saves a fraction ---constant or random--- of their resources~\cite{sinha2003stochastic, chatterjee2004pareto, chakraborti2000statistical, patriarca2004statistical, iglesias2003wealth, iglesias2004correlation, scafetta2002pareto}. Specifically, each agent $i$ is characterized by a wealth $w_i$ and a risk-aversion factor $\beta_i$. 
During the wealth exchange between agents $i$ and $j$,  assuming that $i$ wins, we have $$w_i^* = w_i + dw \>\>\> \text{and} \>\>\> w_j^* = w_j - dw,$$
where $w_{i(j)}^*$ is the wealth of the agent $i(j)$ after the exchange. There are different ways of defining the quantity $dw$ transferred from the loser to the winner, but mostly two are considered: the {\it fair} (or {\it yard-sale}) rule and the {\it loser} rule~\cite{hayes2002, caon2007unfair, cardoso2020wealth}. The first one states that $dw=\min[(1-\beta_{i})w_{i}(t);(1-\beta_{j})w_{j}(t)]$, while in the second we have $dw = (1-\beta_j)w_j(t)$. The first rule is called {\it fair} because the amount of wealth exchanged is the minimum of the quantities put at stake by the two agents and it is the same regardless of who wins, so the richest agent accepts to risk part of its wealth~\cite{hayes2002}. In the {\it loser} rule, on the other hand, the winner receives the value risked by the loser; thus, as the richest agent most probably puts a larger amount at stake, this kind of exchange is only likely in situations where agents do not know the wealth of the others~\cite{sinha2003stochastic}.

Numerical~\cite{iglesias2004correlation, caon2007unfair} and
analytical~\cite{moukarzel2007wealth} results with the {\it
  fair} rule model, or some variations of it, point out to the {\it condensation} fate, {\em i.e.} a continuous
concentration of all available wealth in just one or a few agents,
leading to an absorbing state where no more wealth is exchanged, a
kind of ``thermal death'' of trade~\cite{iglesias2012entropy}. These results seem to describe somehow the present path of the world economy, that
experiences a continuous grow of inequality~\cite{piketty2014capital}.

Different modifications have been introduced in the models to overcome condensation. For example, increasing the probability
of favoring the poorest agent in a
transaction~\cite{iglesias2004correlation, scafetta2002pareto}, or introducing a taxation mechanism~\cite{li2019affine,
  bustos2016wealth}, where periodically all agents pay taxes and the amount collected is then in someway divided among the agents. Here, we study and generalize these two approaches.

In Section~\ref{social} we study the former approach, using a rule suggested by Scafetta~\cite{caon2007unfair,scafetta2002pareto,laguna2005economic, fuentes2006living}, where, in the exchange between the agents $i$ and $j$, the probability of favoring the poorest partner is given by:
\begin{equation}
\label{eq:sca}
p=\frac{1}{2}+f\times\frac{|w_{i}(t)-w_{j}(t)|}{w_{i}(t)+w_{j}(t)},
\end{equation}
where $f$ is a factor that we call {\it social protection} factor, which goes from $0$ (equal probability for both agents) to $1/2$ 
(highest probability of favoring the poorest agent). 
In each interaction the poorest agent has probability $p$ of earn a quantity $dw$, 
whereas the richest one has probability $1-p$. 
It is evident that the higher the difference of wealth in a given pair of agents, the higher the influence of $f$ in the probability. 
For that reason we think $f$ could be an indicator of the degree of government regulation or control of the economy of a country.

However, although with this approach we avoid condensation, the factor $f$, in Eq.~\ref{eq:sca}, can not be transferred easily to quantitative economic measures in real scenarios.
Besides, it can not account for the low values of the Gini index of some countries such as the Scandinavian countries, for example.
Then, in Section~\ref{taxes} we study how different types of tax collection and redistribution affect the distribution of wealth. 

It is important to mention that hereafter  wealth ($w_i$) and risk aversion factor ($\beta_i$) are distributed uniformly in the interval $[0, 1]$ as initial conditions. While $w_i$'s evolve in time as part of the dynamics, the values of $\beta_i$'s are fixed.

\subsection{Inequality}\label{ineq}
To measure inequality, we use three indicators: the share of the wealth of the wealthiest $1\%$ of the agents, the share of the wealth of the richest $10\%$ of the agents (or, reciprocally, the share of wealth in hands of the poorest $90\%$ of the agents), and the Gini index. 

The Gini index is a measure created in 1912 by Italian statistician Conrado Gini; it is usually used to measure the inequality of a distribution of income or wealth. It is calculated as a ratio  of the areas in the Lorenz curve diagram (see Fig.~\ref{fig:lorenz}); if the area between the line of perfect equality and the Lorenz curve is $a$, and the area below the Lorenz curve is $b$, then the Gini coefficient is $a/(a + b)$~\cite{gini1921measurement}. In an operational way we evaluate the Gini index as~\cite{sen1997economic}:
\begin{equation}
G(t) = \frac{1}{2}\frac{\sum_{i,j}|w_i(t) - w_j(t)|}{N \sum_i w_i}.
\end{equation}
The Gini index varies between $0$, which corresponds to perfect equality ({\em i.e.} everyone has the same wealth), and $1$, that corresponds to the extreme inequality where only one agent possesses all the wealth.

\section{Social Protection}\label{social}

\begin{figure}[!htb]
	\centering
	\includegraphics[width=0.7\textwidth]{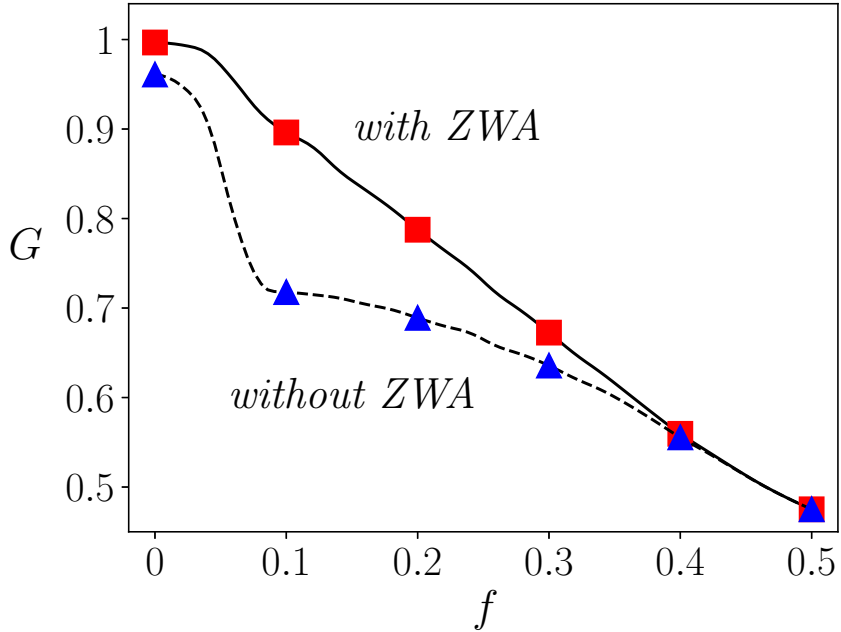}
	\caption{Gini index at equilibrium as a function of the social protection factor, $f$; symbols are calculated with $N=10^5$ and interpolating curves, with values of $f$ separated by $0.01$, with $N=10^4$. {\em ZWA} stands for zero wealth agents.}
	\label{fig:gini_f} 
\end{figure}

We show in Fig.~\ref{fig:gini_f} the Gini index at equilibrium as a function of the factor $f$~\cite{cardoso2020wealth}. 
Clearly, the lower the social protection factor, the greater the inequality until reaching the extreme case $f = 0$, where the system condenses ($G=1$). A significant contribution to the high inequality for low values of $f$ is the large fraction of agents with zero wealth,~\footnote{The global per-capita wealth in 2014 was less than $~10^{5}$ USD~\cite{lange2018changing}, while in our artificial system the mean wealth is $0.5$. Since USD unit is discretized by the minimum value of $0.01$ USD, by analogy we set the minimum unit as $10^{-7}$. Therefore, we consider the wealth of an agent $i$ equal to zero if $w_i < 10^{-7}$.} which is greater than $60\%$ for $f = 0.1$. 
This is evident looking at the comparatively lower Gini curve when zero wealth agents are excluded. It is worth to mention that all results are averages over $10^3$ samples, and three different sizes of the system:  $N $ equal to  $10^3, 10^4$ and $10^5$. As the obtained results are almost independent of the size of the system, we have plotted just the outcome for $N=10^5$ and $N=10^4$.

\begin{figure}
	\centering
	\includegraphics[width=0.45\textwidth]{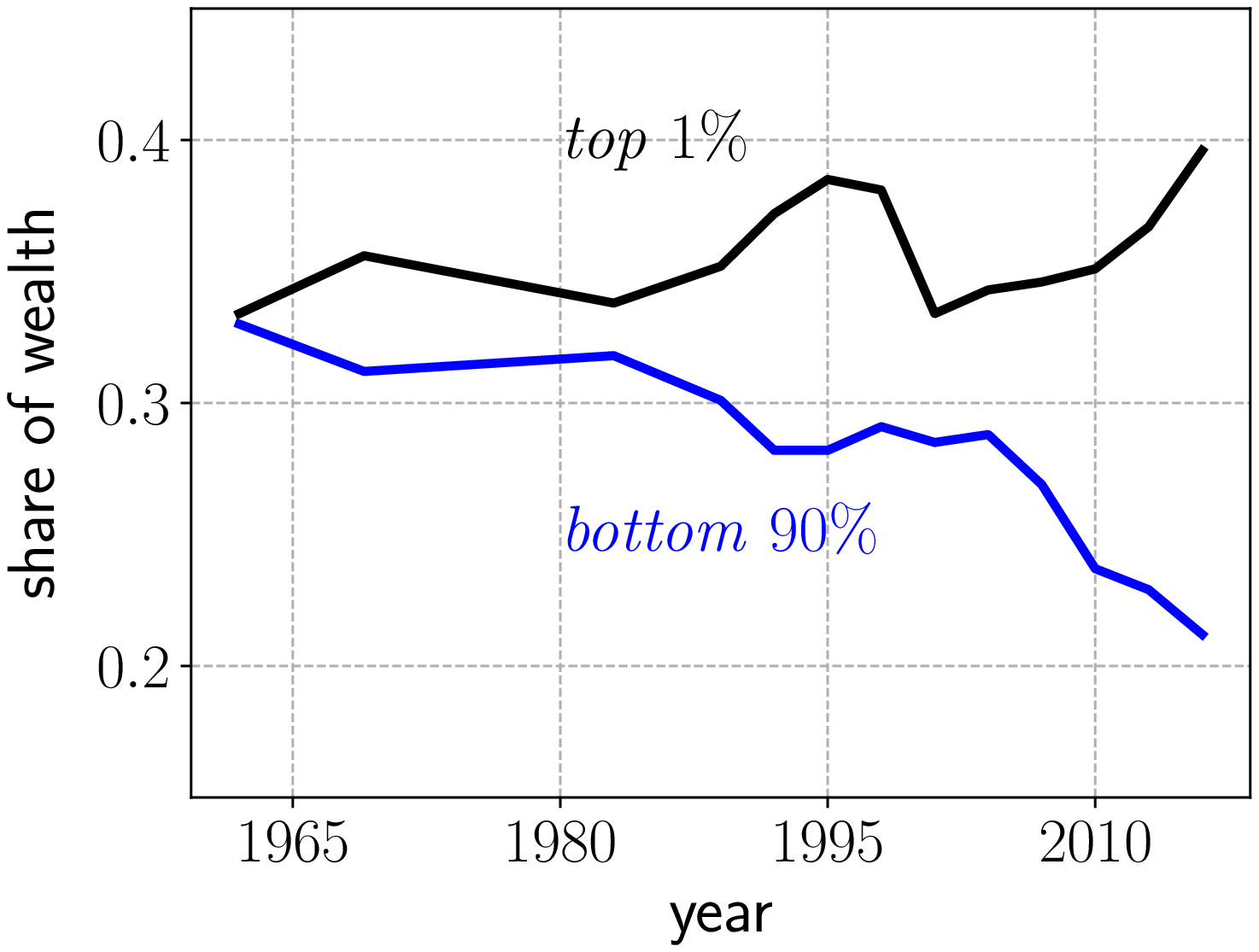}
	\includegraphics[width=0.45\textwidth]{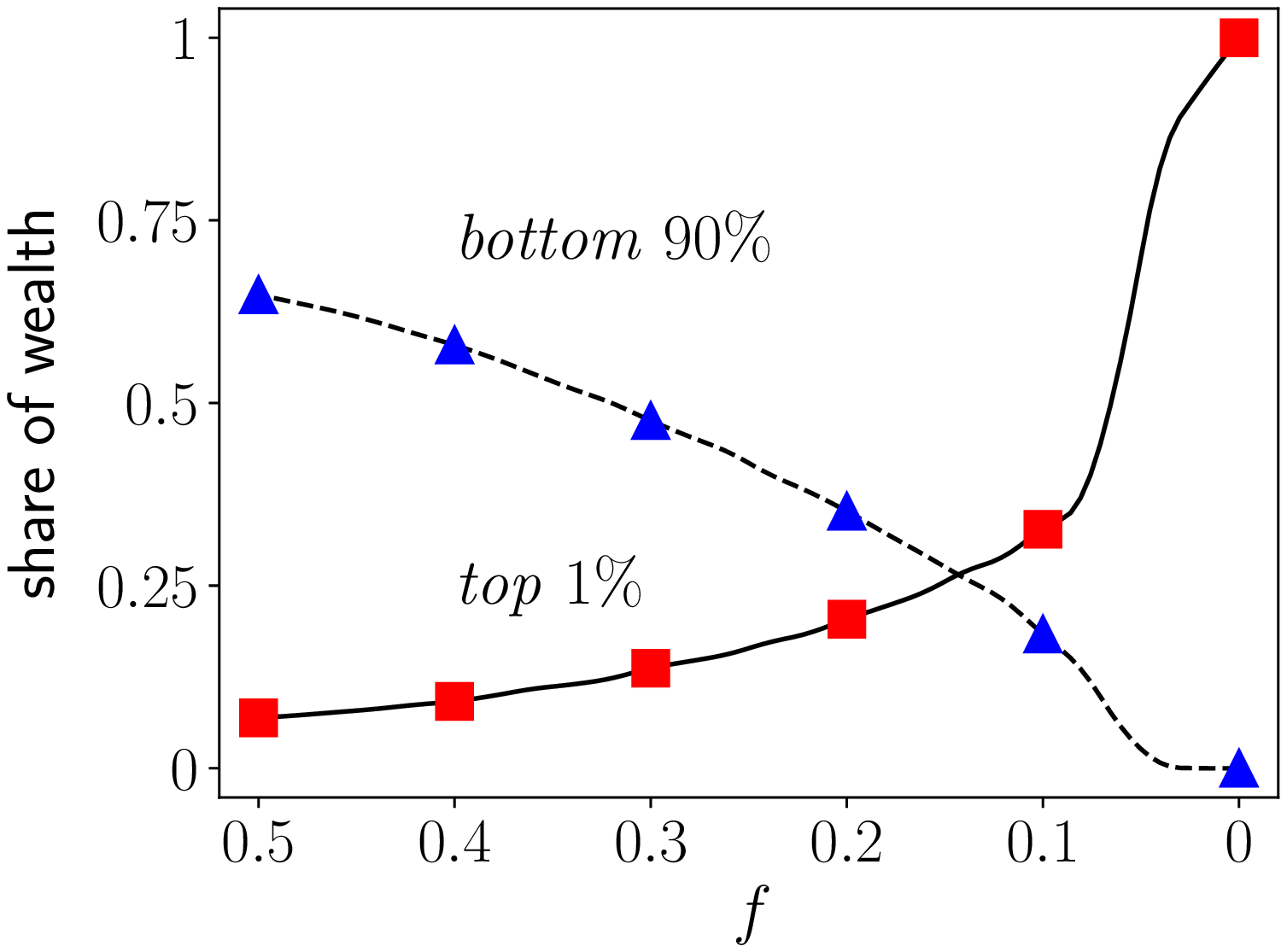}

	\caption{Fraction of wealth in hands of the top 1\% and bottom 90\%. Left: Evolution in the United States; data collected from Wolff~\cite{wolff2017household}. Right: results for the {\it fair rule} in terms of the protection parameter $f$; symbols are calculated with $N=10^5$ and interpolating curves, with values of $f$ separated by $0.01$, with $N=10^4$. Noticed the qualitative parallelism in the two figures for $f<0.15$ which allows to interpret the historical evolution of the US wealth distribution as a temporal slow decrease of a kind of effective $f$ protection factor.} 
	\label{fig:usa}
\end{figure}

A similar relation between social protection and inequality is observed in Fig~\ref{fig:usa} (right), were we show the share of wealth of the upper $1\%$ and the bottom $90\%$ of the agents. Empirically, Fig.~\ref{fig:usa} (left) shows the evolution of the concentration of wealth in the hands of the upper $1\%$ of the US society during the last fifty years, accompanied by the continuous declining of the wealth of the bottom $90\%$. We can interpret the growth of US inequality as the decrease in time of some effective $f$ factor. 

\begin{figure}[!htb]
	\centering
	\includegraphics[width=0.7\textwidth]{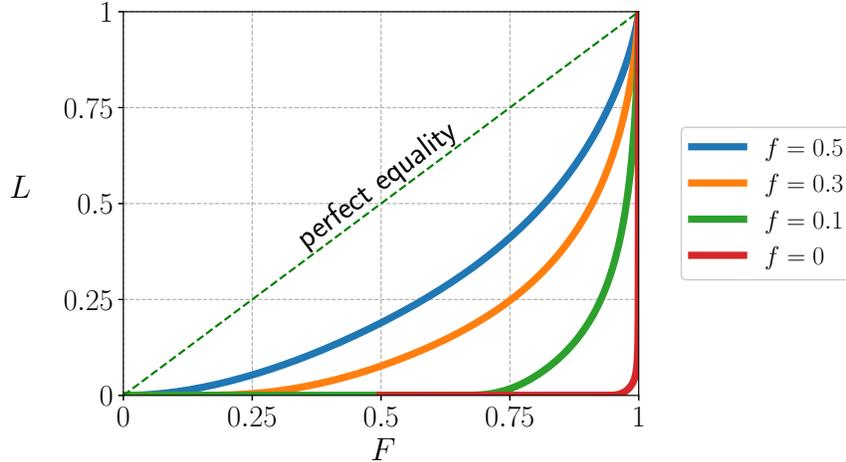}
	\caption{Lorenz curves for the {\it fair} rule  with $N = 10^5$ agents and for different values of the parameter $f$. $L$ is the fraction of wealth that the fraction $F$ of agents has.} 
	\label{fig:lorenz}
\end{figure}

As complementary information, we represent in Fig.~\ref{fig:lorenz} the distribution of wealth by drawing the Lorenz curves for some selected values of $f$.  
One point in a Lorenz curve tells us the fraction of total wealth that a fraction of the population possess. The effect of reducing the protection factor $f$ is very clear, producing curves that depart from the perfect equality line to the almost perfect inequality line in the extreme $f=0$ case.

\section{Taxes}\label{taxes}

In this section, we present results with a simple flat tax system on the wealth. In the simulation, the tax collection works as follows: at each Monte Carlo Step, all agents pay a fraction $\lambda$ of its wealth as taxes~\footnote{The wealth tax is equivalent to taxes on property or fortune, much less widespread than the income tax. However, we choose the former system because it is much effective in reverting inequality.}. Then, the total amount collected is redistributed among the agents, according to two different criteria: universal and targeted. Different kinds of taxes have been studied in ref.~\cite{de2017rich}.

\subsection{Universal Redistribution}\label{univ}
Universal redistribution is the simplest case, where the total amount collected as taxes is equally distributed among all agents, irrespectively of their wealth.  Similar taxation mechanism have already been proposed~\cite{bustos2016wealth, li2019affine}, but assuming that $\beta$ values are very close to $1$, and in the {\it small transaction limit} approximation. Despite this difference, our results are qualitatively analogous.

\begin{figure}[!htb]
	\centering
	\includegraphics[width=0.7\textwidth]{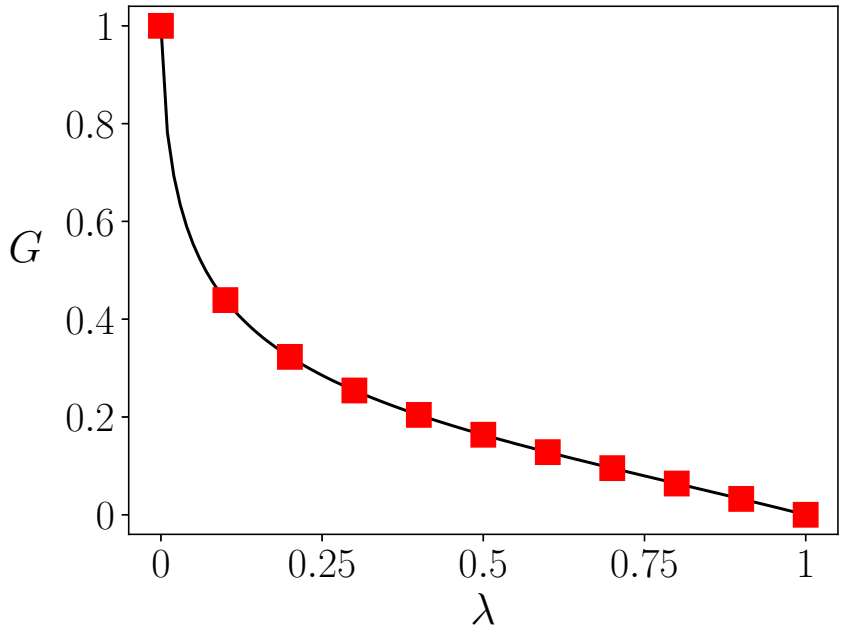}
	\caption{Equilibrium Gini index as a function of $\lambda$, the tax index on fortune.  Lines correspond to simulations with $N=10^4$ agents and symbols, with $N=10^5$.}
	\label{fig:gini_lambda}
\end{figure}

\begin{figure}[!htb]
	\centering
	\includegraphics[width=0.7\textwidth]{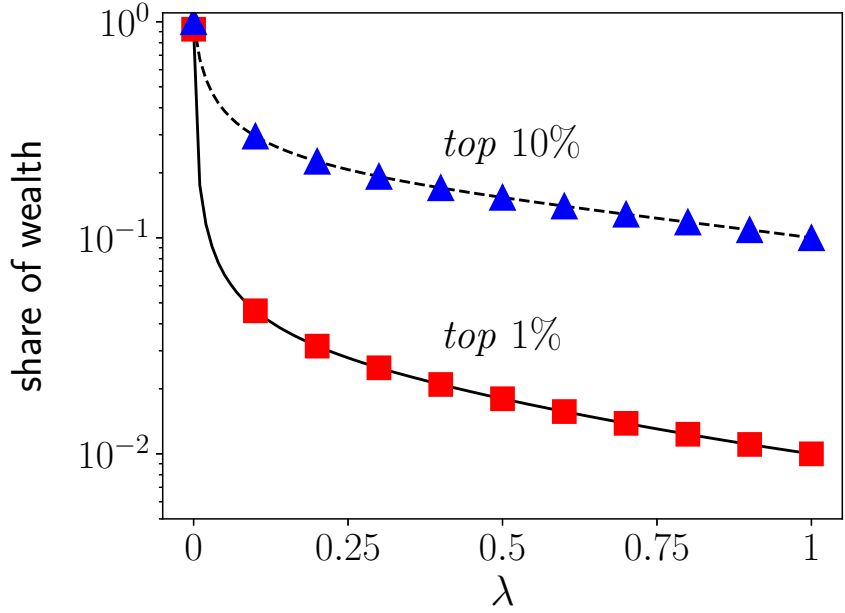}
	\caption{Share of wealth of the top $1\%$ and $10\%$ of the population as a function of $\lambda$, the tax index on fortune.  Lines correspond to simulations with $N=10^4$ agents and symbols, with $N=10^5$.}
	\label{fig:share_lambda}
\end{figure}

We show in Fig.~\ref{fig:gini_lambda} and Fig.~\ref{fig:share_lambda}, respectively, the Gini index and the share of wealth of the top $1\%$ and top $10\%$ of the population as a function of $\lambda$, the tax percentage. We can observe that the higher the tax rate, the lower the inequality, as expected. Differently from the social protection factor $f$, the taxation mechanism can greatly reduce inequality. Moreover, by means of the tax index $\lambda$, it is possible to obtain all the possible values of the Gini index, from the maximum inequality when there are no taxes ($\lambda=0$) to the trivial complete egalitarian society $G=0$, if $\lambda=1$.
However, in real societies one does not expect the value of $\lambda$ to go beyond $0.25$. 

\subsection{Targeted Redistribution}\label{redis}
\begin{figure}[!htb]
	\centering
	\includegraphics[width=0.7\textwidth]{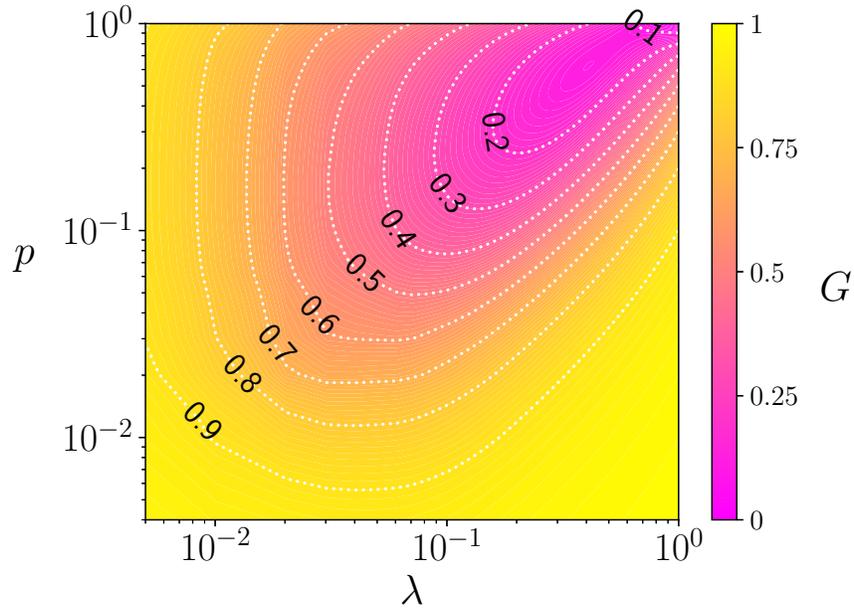}
	\caption{Equilibrium Gini index as a function of $\lambda$  and $p$ (the bottom fraction of agents). Results for $N = 10^4$ agents.}
	\label{fig:gini_lambda_p}
\end{figure}

In the targeted case, the total amount collected as taxes is distributed among the $p$ poorest fraction of the population (the targeted population). The universal case is recovered when $p = 1$. We show in Fig.~\ref{fig:gini_lambda_p} the Gini index as a function of both $\lambda$ and $p$. We remark that, by restricting the allocations to less that 1\% of the people ($p \leq 10^{-2}$), as it is the case in most of governmental projets to help unemployed and/or very poor people, the effect, in terms of Gini coefficient, is almost unperceptible. Instead, it is necessary to expand assistance to at least 10\% of the poorest people to achieve a real effect of reducing inequality.
The figure also shows that there is an optimum value of $p = p^*$ that minimize the inequality for each value of the tax rate $\lambda$, so there is an interesting non-trivial relation between $\lambda$ and $p$ in this case.

\begin{figure}[!htb]
	\centering
	\includegraphics[width=0.7\textwidth]{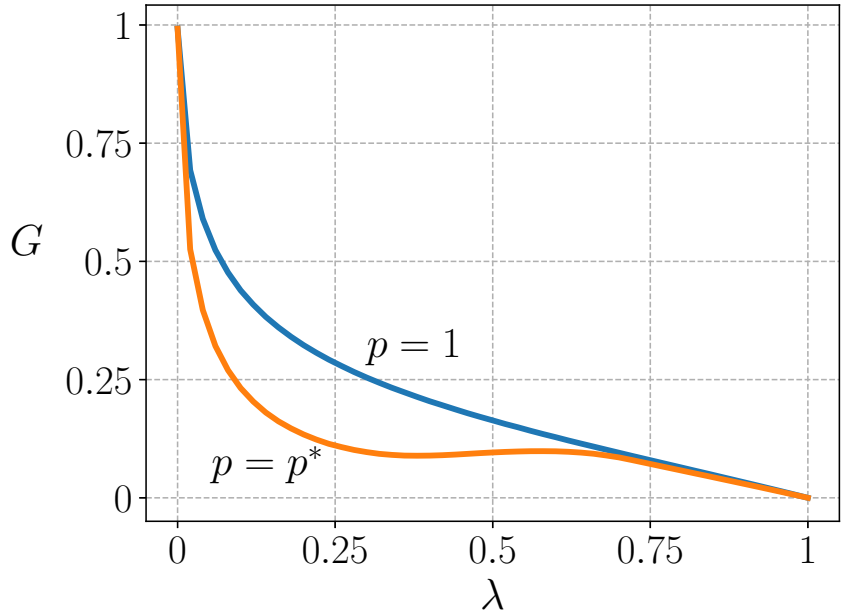}
	\caption{Equilibrium Gini index as a function of  $\lambda$ (tax index on wealth), for $p=1$ (universal case) and $p=p^*$ (optimal targeted case). Results for $N = 10^4$ agents.}
	\label{fig:gini_lambda_opt}
\end{figure}

Finally, in Fig.~\ref{fig:gini_lambda_opt} we compare the Gini index as a function of $\lambda$, for $p=1$ (universal case) and $p=p^*$ (optimal targeted case). We remark that, for intermediary values of $\lambda$, particularly for $\lambda \approx 0.35$ the regulatory mechanism of helping just a fraction of the population is more efficient to strongly reduce inequality.

\section{Discussion and Conclusion}\label{disc}

Economic inequalities are continuously increasing worldwide, with the exception of a few countries. It is a plague of the past and present centuries that threatens the economic sustainability all around the world.
Inequalities arise in a very similar way to the condensation tendency of the models reviewed here. The ideal free market without regulations, which seems to be fair because it gives everyone the same opportunities, it is not. It is an illusion that fails because of the multiplicative factor behind, which increases the wealth of the first favored by random fluctuations, while the rest of the population goes irremediably to breakdown.
We have shown that even strong regulations, while avoiding condensation, are not enough to lead to a low Gini index.
On the other hand, we have demonstrated that taxes are a possible way to get to any level of inequality, from the highest to the utopic minimum. A practical evidence is the case of Scandinavian countries (where the Gini coefficient is of the order of $G=0.25$) which exhibit both high taxes and low inequality.

The model here presented is certainly a very simplified one; for example, taxes do not play just a redistributive objective, but also are used by governments to finance public administration and to assure the infrastructure needed to economic development. Nonetheless, in spite on its simplicity the present model describes a very strong redistribution mechanism, that coincides with some public policies, like the ``bolsa-familia'' (family fellowship, allocate to very poor family groups in Brazil), or the small loans to jobless people in order to develop their own business. Even at global level, and with taxes proportional to the possessions of each agent, the contribution of each one could be small but results in a low Gini coefficient, even lower than the one observed in the most egalitarian countries, like Denmark or Japan . 
Is has to be also noticed that the universal redistribution tax system is much easier to administrate than a targeted redistribution system.
Even if the tax on fortune seems a very reasonable mechanism to reduce inequalities, we are presently working in the effect of other kind of taxes, like tax on the profits, revenues and consumption. Also, more sofisticated redistributive policies are also investigated. Results will be communicated soon.  

We conclude that the yard-sale model, in spite of its simplicity, is able to reproduce some properties of modern economies, particularly the tendency to continuously increase inequality in the absence of regulations. We have also demonstrated that an  effective redestributive mechanism, provided by a policy of taxing great fortunes and incomes, can greatly reduce inequality. 

\section*{Acknowledgements}

This study was partially financed by the Coordenação de Aperfeiçoamento de Pessoal de Nível Superior - Brasil (CAPES) -
Finance Code 001.
BFC acknowledges Brazilian agency Conselho Nacional de Desenvolvimento Científico e Tecnológico (CNPq) for scholarship.    
JRI acknowledges Brazilian agency Conselho Nacional de Desenvolvimento Científico e Tecnológico (CNPq) for 
support. SG acknowledges CAPES for support, under fellowship  \#003/2019 - PROPG - PRINT/UFRGS, and URPP Social Networks and University of Zürich for hospitality.

Last but not least, we thank D. Chialvo for organizing the Complexity Weekend, for asking and agreeing that an article of this sort should be written, and for editing and getting this volume to the press. 

%%-----------------------------
%%      your bibliography

\bibliography{ref}

\begin{thebibliography}{32}%
\makeatletter
\providecommand \@ifxundefined [1]{%
 \@ifx{#1\undefined}
}%
\providecommand \@ifnum [1]{%
 \ifnum #1\expandafter \@firstoftwo
 \else \expandafter \@secondoftwo
 \fi
}%
\providecommand \@ifx [1]{%
 \ifx #1\expandafter \@firstoftwo
 \else \expandafter \@secondoftwo
 \fi
}%
\providecommand \natexlab [1]{#1}%
\providecommand \enquote  [1]{``#1''}%
\providecommand \bibnamefont  [1]{#1}%
\providecommand \bibfnamefont [1]{#1}%
\providecommand \citenamefont [1]{#1}%
\providecommand \href@noop [0]{\@secondoftwo}%
\providecommand \href [0]{\begingroup \@sanitize@url \@href}%
\providecommand \@href[1]{\@@startlink{#1}\@@href}%
\providecommand \@@href[1]{\endgroup#1\@@endlink}%
\providecommand \@sanitize@url [0]{\catcode `\\12\catcode `\$12\catcode
  `\&12\catcode `\#12\catcode `\^12\catcode `\_12\catcode `\%12\relax}%
\providecommand \@@startlink[1]{}%
\providecommand \@@endlink[0]{}%
\providecommand \url  [0]{\begingroup\@sanitize@url \@url }%
\providecommand \@url [1]{\endgroup\@href {#1}{\urlprefix }}%
\providecommand \urlprefix  [0]{URL }%
\providecommand \Eprint [0]{\href }%
\providecommand \doibase [0]{http://dx.doi.org/}%
\providecommand \selectlanguage [0]{\@gobble}%
\providecommand \bibinfo  [0]{\@secondoftwo}%
\providecommand \bibfield  [0]{\@secondoftwo}%
\providecommand \translation [1]{[#1]}%
\providecommand \BibitemOpen [0]{}%
\providecommand \bibitemStop [0]{}%
\providecommand \bibitemNoStop [0]{.\EOS\space}%
\providecommand \EOS [0]{\spacefactor3000\relax}%
\providecommand \BibitemShut  [1]{\csname bibitem#1\endcsname}%
\let\auto@bib@innerbib\@empty
%</preamble>
\bibitem [{\citenamefont {Wolff}(2017)}]{wolff2017household}%
  \BibitemOpen
  \bibfield  {author} {\bibinfo {author} {\bibfnamefont {E.~N.}\ \bibnamefont
  {Wolff}},\ }\href@noop {} {\emph {\bibinfo {title} {Household Wealth Trends
  in the United States, 1962 to 2016: Has Middle Class Wealth Recovered?}}},\
  \bibinfo {type} {Tech. Rep.}\ (\bibinfo  {institution} {National Bureau of
  Economic Research},\ \bibinfo {year} {2017})\BibitemShut {NoStop}%
\bibitem [{\citenamefont {Piketty}(2014)}]{piketty2014capital}%
  \BibitemOpen
  \bibfield  {author} {\bibinfo {author} {\bibfnamefont {T.}~\bibnamefont
  {Piketty}},\ }\href@noop {} {\emph {\bibinfo {title} {Capital in the 21st
  Century}}}\ (\bibinfo  {publisher} {Harvard University Press Cambridge, MA},\
  \bibinfo {year} {2014})\BibitemShut {NoStop}%
\bibitem [{\citenamefont {Pareto}(1897)}]{pareto1897cours}%
  \BibitemOpen
  \bibfield  {author} {\bibinfo {author} {\bibfnamefont {V.}~\bibnamefont
  {Pareto}},\ }\href@noop {} {\bibfield  {journal} {\bibinfo  {journal} {Rouge,
  Lausanne}\ }\textbf {\bibinfo {volume} {2}} (\bibinfo {year}
  {1897})}\BibitemShut {NoStop}%
\bibitem [{\citenamefont {Stanley}\ and\ \citenamefont
  {Mantegna}(2000)}]{stanley2000introduction}%
  \BibitemOpen
  \bibfield  {author} {\bibinfo {author} {\bibfnamefont {H.~E.}\ \bibnamefont
  {Stanley}}\ and\ \bibinfo {author} {\bibfnamefont {R.~N.}\ \bibnamefont
  {Mantegna}},\ }\href@noop {} {\emph {\bibinfo {title} {An introduction to
  econophysics}}}\ (\bibinfo  {publisher} {Cambridge University Press,
  Cambridge},\ \bibinfo {year} {2000})\BibitemShut {NoStop}%
\bibitem [{\citenamefont {Levy}\ and\ \citenamefont
  {Levy}(2003)}]{levy2003investment}%
  \BibitemOpen
  \bibfield  {author} {\bibinfo {author} {\bibfnamefont {M.}~\bibnamefont
  {Levy}}\ and\ \bibinfo {author} {\bibfnamefont {H.}~\bibnamefont {Levy}},\
  }\href@noop {} {\bibfield  {journal} {\bibinfo  {journal} {Review of
  Economics and Statistics}\ }\textbf {\bibinfo {volume} {85}},\ \bibinfo
  {pages} {709} (\bibinfo {year} {2003})}\BibitemShut {NoStop}%
\bibitem [{\citenamefont {Richmond}\ \emph {et~al.}(2006)\citenamefont
  {Richmond}, \citenamefont {Hutzler}, \citenamefont {Coelho},\ and\
  \citenamefont {Repetowicz}}]{richmond2006review}%
  \BibitemOpen
  \bibfield  {author} {\bibinfo {author} {\bibfnamefont {P.}~\bibnamefont
  {Richmond}}, \bibinfo {author} {\bibfnamefont {S.}~\bibnamefont {Hutzler}},
  \bibinfo {author} {\bibfnamefont {R.}~\bibnamefont {Coelho}}, \ and\ \bibinfo
  {author} {\bibfnamefont {P.}~\bibnamefont {Repetowicz}},\ }\href@noop {}
  {\emph {\bibinfo {title} {A review of empirical studies and models of income
  distributions in society}}}\ (\bibinfo  {publisher} {Wiley-VCH: Berlin,
  Germany},\ \bibinfo {year} {2006})\BibitemShut {NoStop}%
\bibitem [{\citenamefont {Sinha}(2006)}]{sinha2006evidence}%
  \BibitemOpen
  \bibfield  {author} {\bibinfo {author} {\bibfnamefont {S.}~\bibnamefont
  {Sinha}},\ }\href@noop {} {\bibfield  {journal} {\bibinfo  {journal} {Physica
  A: Statistical Mechanics and its Applications}\ }\textbf {\bibinfo {volume}
  {359}},\ \bibinfo {pages} {555} (\bibinfo {year} {2006})}\BibitemShut
  {NoStop}%
\bibitem [{\citenamefont {Klass}\ \emph {et~al.}(2007)\citenamefont {Klass},
  \citenamefont {Biham}, \citenamefont {Levy}, \citenamefont {Malcai},\ and\
  \citenamefont {Solomon}}]{klass2007forbes}%
  \BibitemOpen
  \bibfield  {author} {\bibinfo {author} {\bibfnamefont {O.~S.}\ \bibnamefont
  {Klass}}, \bibinfo {author} {\bibfnamefont {O.}~\bibnamefont {Biham}},
  \bibinfo {author} {\bibfnamefont {M.}~\bibnamefont {Levy}}, \bibinfo {author}
  {\bibfnamefont {O.}~\bibnamefont {Malcai}}, \ and\ \bibinfo {author}
  {\bibfnamefont {S.}~\bibnamefont {Solomon}},\ }\href@noop {} {\bibfield
  {journal} {\bibinfo  {journal} {The European Physical Journal B}\ }\textbf
  {\bibinfo {volume} {55}},\ \bibinfo {pages} {143} (\bibinfo {year}
  {2007})}\BibitemShut {NoStop}%
\bibitem [{\citenamefont {Dr{\u{a}}gulescu}\ and\ \citenamefont
  {Yakovenko}(2001)}]{druagulescu2001exponential}%
  \BibitemOpen
  \bibfield  {author} {\bibinfo {author} {\bibfnamefont {A.}~\bibnamefont
  {Dr{\u{a}}gulescu}}\ and\ \bibinfo {author} {\bibfnamefont {V.~M.}\
  \bibnamefont {Yakovenko}},\ }\href@noop {} {\bibfield  {journal} {\bibinfo
  {journal} {Physica A: Statistical Mechanics and its Applications}\ }\textbf
  {\bibinfo {volume} {299}},\ \bibinfo {pages} {213} (\bibinfo {year}
  {2001})}\BibitemShut {NoStop}%
\bibitem [{IRP()}]{IRPF}%
  \BibitemOpen
  \href@noop {} {\enquote {\bibinfo {title} {Iprf grandes números},}\
  }\bibinfo {howpublished}
  {\url{http://receita.economia.gov.br/dados/receitadata/estudos-e-tributarios-e-aduaneiros/estudos-e-estatisticas/11-08-2014-grandes-numeros-dirpf/grandes-numeros-dirpf-capa}},\
  \bibinfo {note} {accessed: 2020-01-18}\BibitemShut {NoStop}%
\bibitem [{IBG()}]{IBGE}%
  \BibitemOpen
  \href@noop {} {\enquote {\bibinfo {title} {Ibge pib municípios},}\ }\bibinfo
  {howpublished} {\url{ftp://ftp.ibge.gov.br/Pib_Municipios/2015/}},\ \bibinfo
  {note} {accessed: 2020-01-18}\BibitemShut {NoStop}%
\bibitem [{\citenamefont {Caon}\ \emph {et~al.}(2007)\citenamefont {Caon},
  \citenamefont {Gon{\c{c}}alves},\ and\ \citenamefont
  {Iglesias}}]{caon2007unfair}%
  \BibitemOpen
  \bibfield  {author} {\bibinfo {author} {\bibfnamefont {G.}~\bibnamefont
  {Caon}}, \bibinfo {author} {\bibfnamefont {S.}~\bibnamefont
  {Gon{\c{c}}alves}}, \ and\ \bibinfo {author} {\bibfnamefont {J.}~\bibnamefont
  {Iglesias}},\ }\href@noop {} {\bibfield  {journal} {\bibinfo  {journal} {The
  European Physical Journal Special Topics}\ }\textbf {\bibinfo {volume}
  {143}},\ \bibinfo {pages} {69} (\bibinfo {year} {2007})}\BibitemShut
  {NoStop}%
\bibitem [{\citenamefont {Li}\ \emph {et~al.}(2019)\citenamefont {Li},
  \citenamefont {Boghosian},\ and\ \citenamefont {Li}}]{li2019affine}%
  \BibitemOpen
  \bibfield  {author} {\bibinfo {author} {\bibfnamefont {J.}~\bibnamefont
  {Li}}, \bibinfo {author} {\bibfnamefont {B.~M.}\ \bibnamefont {Boghosian}}, \
  and\ \bibinfo {author} {\bibfnamefont {C.}~\bibnamefont {Li}},\ }\href@noop
  {} {\bibfield  {journal} {\bibinfo  {journal} {Physica A: Statistical
  Mechanics and its Applications}\ }\textbf {\bibinfo {volume} {516}},\
  \bibinfo {pages} {423} (\bibinfo {year} {2019})}\BibitemShut {NoStop}%
\bibitem [{\citenamefont {Yakovenko}\ and\ \citenamefont {{ROSSER
  JR}}(2009)}]{yakovenko2009colloquium}%
  \BibitemOpen
  \bibfield  {author} {\bibinfo {author} {\bibfnamefont {V.~M.}\ \bibnamefont
  {Yakovenko}}\ and\ \bibinfo {author} {\bibfnamefont {J.~B.}\ \bibnamefont
  {{ROSSER JR}}},\ }\href@noop {} {\bibfield  {journal} {\bibinfo  {journal}
  {Reviews of Modern Physics}\ }\textbf {\bibinfo {volume} {81}},\ \bibinfo
  {pages} {1703} (\bibinfo {year} {2009})}\BibitemShut {NoStop}%
\bibitem [{\citenamefont {Sinha}(2003)}]{sinha2003stochastic}%
  \BibitemOpen
  \bibfield  {author} {\bibinfo {author} {\bibfnamefont {S.}~\bibnamefont
  {Sinha}},\ }\href@noop {} {\bibfield  {journal} {\bibinfo  {journal} {Physica
  Scripta}\ }\textbf {\bibinfo {volume} {2003}},\ \bibinfo {pages} {59}
  (\bibinfo {year} {2003})}\BibitemShut {NoStop}%
\bibitem [{\citenamefont {Chatterjee}\ \emph {et~al.}(2004)\citenamefont
  {Chatterjee}, \citenamefont {Chakrabarti},\ and\ \citenamefont
  {Manna}}]{chatterjee2004pareto}%
  \BibitemOpen
  \bibfield  {author} {\bibinfo {author} {\bibfnamefont {A.}~\bibnamefont
  {Chatterjee}}, \bibinfo {author} {\bibfnamefont {B.~K.}\ \bibnamefont
  {Chakrabarti}}, \ and\ \bibinfo {author} {\bibfnamefont {S.}~\bibnamefont
  {Manna}},\ }\href@noop {} {\bibfield  {journal} {\bibinfo  {journal} {Physica
  A: Statistical Mechanics and its Applications}\ }\textbf {\bibinfo {volume}
  {335}},\ \bibinfo {pages} {155} (\bibinfo {year} {2004})}\BibitemShut
  {NoStop}%
\bibitem [{\citenamefont {Chakraborti}\ and\ \citenamefont
  {Chakrabarti}(2000)}]{chakraborti2000statistical}%
  \BibitemOpen
  \bibfield  {author} {\bibinfo {author} {\bibfnamefont {A.}~\bibnamefont
  {Chakraborti}}\ and\ \bibinfo {author} {\bibfnamefont {B.~K.}\ \bibnamefont
  {Chakrabarti}},\ }\href@noop {} {\bibfield  {journal} {\bibinfo  {journal}
  {The European Physical Journal B-Condensed Matter and Complex Systems}\
  }\textbf {\bibinfo {volume} {17}},\ \bibinfo {pages} {167} (\bibinfo {year}
  {2000})}\BibitemShut {NoStop}%
\bibitem [{\citenamefont {Patriarca}\ \emph {et~al.}(2004)\citenamefont
  {Patriarca}, \citenamefont {Chakraborti},\ and\ \citenamefont
  {Kaski}}]{patriarca2004statistical}%
  \BibitemOpen
  \bibfield  {author} {\bibinfo {author} {\bibfnamefont {M.}~\bibnamefont
  {Patriarca}}, \bibinfo {author} {\bibfnamefont {A.}~\bibnamefont
  {Chakraborti}}, \ and\ \bibinfo {author} {\bibfnamefont {K.}~\bibnamefont
  {Kaski}},\ }\href@noop {} {\bibfield  {journal} {\bibinfo  {journal}
  {Physical Review E}\ }\textbf {\bibinfo {volume} {70}},\ \bibinfo {pages}
  {016104} (\bibinfo {year} {2004})}\BibitemShut {NoStop}%
\bibitem [{\citenamefont {Iglesias}\ \emph {et~al.}(2003)\citenamefont
  {Iglesias}, \citenamefont {Gon{\c{c}}alves}, \citenamefont {Pianegonda},
  \citenamefont {Vega},\ and\ \citenamefont {Abramson}}]{iglesias2003wealth}%
  \BibitemOpen
  \bibfield  {author} {\bibinfo {author} {\bibfnamefont {J.~R.}\ \bibnamefont
  {Iglesias}}, \bibinfo {author} {\bibfnamefont {S.}~\bibnamefont
  {Gon{\c{c}}alves}}, \bibinfo {author} {\bibfnamefont {S.}~\bibnamefont
  {Pianegonda}}, \bibinfo {author} {\bibfnamefont {J.~L.}\ \bibnamefont
  {Vega}}, \ and\ \bibinfo {author} {\bibfnamefont {G.}~\bibnamefont
  {Abramson}},\ }\href@noop {} {\bibfield  {journal} {\bibinfo  {journal}
  {Physica A: Statistical Mechanics and its Applications}\ }\textbf {\bibinfo
  {volume} {327}},\ \bibinfo {pages} {12} (\bibinfo {year} {2003})}\BibitemShut
  {NoStop}%
\bibitem [{\citenamefont {Iglesias}\ \emph {et~al.}(2004)\citenamefont
  {Iglesias}, \citenamefont {Gon{\c{c}}alves}, \citenamefont {Abramson},\ and\
  \citenamefont {Vega}}]{iglesias2004correlation}%
  \BibitemOpen
  \bibfield  {author} {\bibinfo {author} {\bibfnamefont {J.~R.}\ \bibnamefont
  {Iglesias}}, \bibinfo {author} {\bibfnamefont {S.}~\bibnamefont
  {Gon{\c{c}}alves}}, \bibinfo {author} {\bibfnamefont {G.}~\bibnamefont
  {Abramson}}, \ and\ \bibinfo {author} {\bibfnamefont {J.~L.}\ \bibnamefont
  {Vega}},\ }\href@noop {} {\bibfield  {journal} {\bibinfo  {journal} {Physica
  A: Statistical Mechanics and its Applications}\ }\textbf {\bibinfo {volume}
  {342}},\ \bibinfo {pages} {186} (\bibinfo {year} {2004})}\BibitemShut
  {NoStop}%
\bibitem [{\citenamefont {Scafetta}\ \emph {et~al.}(2002)\citenamefont
  {Scafetta}, \citenamefont {Picozzi},\ and\ \citenamefont
  {West}}]{scafetta2002pareto}%
  \BibitemOpen
  \bibfield  {author} {\bibinfo {author} {\bibfnamefont {N.}~\bibnamefont
  {Scafetta}}, \bibinfo {author} {\bibfnamefont {S.}~\bibnamefont {Picozzi}}, \
  and\ \bibinfo {author} {\bibfnamefont {B.~J.}\ \bibnamefont {West}},\
  }\href@noop {} {\bibfield  {journal} {\bibinfo  {journal} {arXiv preprint
  cond-mat/0209373}\ } (\bibinfo {year} {2002})}\BibitemShut {NoStop}%
\bibitem [{\citenamefont {Hayes}(2002)}]{hayes2002}%
  \BibitemOpen
  \bibfield  {author} {\bibinfo {author} {\bibfnamefont {B.}~\bibnamefont
  {Hayes}},\ }\href@noop {} {\bibfield  {journal} {\bibinfo  {journal}
  {American Scientis}\ }\textbf {\bibinfo {volume} {90}},\ \bibinfo {pages}
  {400} (\bibinfo {year} {2002})}\BibitemShut {NoStop}%
\bibitem [{\citenamefont {Cardoso}\ \emph {et~al.}(2020)\citenamefont
  {Cardoso}, \citenamefont {Gon{\c{c}}alves},\ and\ \citenamefont
  {Iglesias}}]{cardoso2020wealth}%
  \BibitemOpen
  \bibfield  {author} {\bibinfo {author} {\bibfnamefont {B.-H.~F.}\
  \bibnamefont {Cardoso}}, \bibinfo {author} {\bibfnamefont {S.}~\bibnamefont
  {Gon{\c{c}}alves}}, \ and\ \bibinfo {author} {\bibfnamefont {J.~R.}\
  \bibnamefont {Iglesias}},\ }\href@noop {} {\bibfield  {journal} {\bibinfo
  {journal} {Physica A: Statistical Mechanics and its Applications}\ ,\
  \bibinfo {pages} {124201}} (\bibinfo {year} {2020})}\BibitemShut {NoStop}%
\bibitem [{\citenamefont {Moukarzel}\ \emph {et~al.}(2007)\citenamefont
  {Moukarzel}, \citenamefont {Gon{\c{c}}alves}, \citenamefont {Iglesias},
  \citenamefont {Rodr{\'\i}guez-Achach},\ and\ \citenamefont
  {Huerta-Quintanilla}}]{moukarzel2007wealth}%
  \BibitemOpen
  \bibfield  {author} {\bibinfo {author} {\bibfnamefont {C.~F.}\ \bibnamefont
  {Moukarzel}}, \bibinfo {author} {\bibfnamefont {S.}~\bibnamefont
  {Gon{\c{c}}alves}}, \bibinfo {author} {\bibfnamefont {J.~R.}\ \bibnamefont
  {Iglesias}}, \bibinfo {author} {\bibfnamefont {M.}~\bibnamefont
  {Rodr{\'\i}guez-Achach}}, \ and\ \bibinfo {author} {\bibfnamefont
  {R.}~\bibnamefont {Huerta-Quintanilla}},\ }\href@noop {} {\bibfield
  {journal} {\bibinfo  {journal} {The European Physical Journal Special
  Topics}\ }\textbf {\bibinfo {volume} {143}},\ \bibinfo {pages} {75} (\bibinfo
  {year} {2007})}\BibitemShut {NoStop}%
\bibitem [{\citenamefont {Iglesias}\ and\ \citenamefont
  {De~Almeida}(2012)}]{iglesias2012entropy}%
  \BibitemOpen
  \bibfield  {author} {\bibinfo {author} {\bibfnamefont {J.~R.}\ \bibnamefont
  {Iglesias}}\ and\ \bibinfo {author} {\bibfnamefont {R.~M.~C.}\ \bibnamefont
  {De~Almeida}},\ }\href@noop {} {\bibfield  {journal} {\bibinfo  {journal}
  {The European Physical Journal B}\ }\textbf {\bibinfo {volume} {85}},\
  \bibinfo {pages} {85} (\bibinfo {year} {2012})}\BibitemShut {NoStop}%
\bibitem [{\citenamefont {Bustos-Guajardo}\ and\ \citenamefont
  {Moukarzel}(2016)}]{bustos2016wealth}%
  \BibitemOpen
  \bibfield  {author} {\bibinfo {author} {\bibfnamefont {R.}~\bibnamefont
  {Bustos-Guajardo}}\ and\ \bibinfo {author} {\bibfnamefont {C.~F.}\
  \bibnamefont {Moukarzel}},\ }\href@noop {} {\bibfield  {journal} {\bibinfo
  {journal} {International Journal of Modern Physics C}\ }\textbf {\bibinfo
  {volume} {27}},\ \bibinfo {pages} {1650094} (\bibinfo {year}
  {2016})}\BibitemShut {NoStop}%
\bibitem [{\citenamefont {Laguna}\ \emph {et~al.}(2005)\citenamefont {Laguna},
  \citenamefont {Risau~Gusman},\ and\ \citenamefont
  {Iglesias}}]{laguna2005economic}%
  \BibitemOpen
  \bibfield  {author} {\bibinfo {author} {\bibfnamefont {M.~F.}\ \bibnamefont
  {Laguna}}, \bibinfo {author} {\bibfnamefont {S.}~\bibnamefont
  {Risau~Gusman}}, \ and\ \bibinfo {author} {\bibfnamefont {J.~R.}\
  \bibnamefont {Iglesias}},\ }\href@noop {} {\bibfield  {journal} {\bibinfo
  {journal} {Physica A: Statistical Mechanics and its Applications}\ }\textbf
  {\bibinfo {volume} {356}},\ \bibinfo {pages} {107} (\bibinfo {year}
  {2005})}\BibitemShut {NoStop}%
\bibitem [{\citenamefont {Fuentes}\ \emph {et~al.}(2006)\citenamefont
  {Fuentes}, \citenamefont {Kuperman},\ and\ \citenamefont
  {Iglesias}}]{fuentes2006living}%
  \BibitemOpen
  \bibfield  {author} {\bibinfo {author} {\bibfnamefont {M.~A.}\ \bibnamefont
  {Fuentes}}, \bibinfo {author} {\bibfnamefont {M.}~\bibnamefont {Kuperman}}, \
  and\ \bibinfo {author} {\bibfnamefont {J.~R.}\ \bibnamefont {Iglesias}},\
  }\href@noop {} {\bibfield  {journal} {\bibinfo  {journal} {Physica A:
  Statistical Mechanics and its Applications}\ }\textbf {\bibinfo {volume}
  {371}},\ \bibinfo {pages} {112} (\bibinfo {year} {2006})}\BibitemShut
  {NoStop}%
\bibitem [{\citenamefont {Gini}(1921)}]{gini1921measurement}%
  \BibitemOpen
  \bibfield  {author} {\bibinfo {author} {\bibfnamefont {C.}~\bibnamefont
  {Gini}},\ }\href@noop {} {\bibfield  {journal} {\bibinfo  {journal} {The
  Economic Journal}\ }\textbf {\bibinfo {volume} {31}},\ \bibinfo {pages} {124}
  (\bibinfo {year} {1921})}\BibitemShut {NoStop}%
\bibitem [{\citenamefont {Sen}\ \emph {et~al.}(1997)\citenamefont {Sen},
  \citenamefont {Sen}, \citenamefont {Amartya}, \citenamefont {Foster},
  \citenamefont {Foster} \emph {et~al.}}]{sen1997economic}%
  \BibitemOpen
  \bibfield  {author} {\bibinfo {author} {\bibfnamefont {A.}~\bibnamefont
  {Sen}}, \bibinfo {author} {\bibfnamefont {M.~A.}\ \bibnamefont {Sen}},
  \bibinfo {author} {\bibfnamefont {S.}~\bibnamefont {Amartya}}, \bibinfo
  {author} {\bibfnamefont {J.~E.}\ \bibnamefont {Foster}}, \bibinfo {author}
  {\bibfnamefont {J.~E.}\ \bibnamefont {Foster}},  \emph {et~al.},\ }\href@noop
  {} {\emph {\bibinfo {title} {On economic inequality}}}\ (\bibinfo
  {publisher} {Oxford University Press},\ \bibinfo {year} {1997})\BibitemShut
  {NoStop}%
\bibitem [{\citenamefont {Lange}\ \emph {et~al.}(2018)\citenamefont {Lange},
  \citenamefont {Wodon},\ and\ \citenamefont {Carey}}]{lange2018changing}%
  \BibitemOpen
  \bibfield  {author} {\bibinfo {author} {\bibfnamefont {G.-M.}\ \bibnamefont
  {Lange}}, \bibinfo {author} {\bibfnamefont {Q.}~\bibnamefont {Wodon}}, \ and\
  \bibinfo {author} {\bibfnamefont {K.}~\bibnamefont {Carey}},\ }\href@noop {}
  {\emph {\bibinfo {title} {The changing wealth of nations 2018: Building a
  sustainable future}}}\ (\bibinfo  {publisher} {The World Bank},\ \bibinfo
  {year} {2018})\BibitemShut {NoStop}%
\bibitem [{\citenamefont {de~Oliveira}(2017)}]{de2017rich}%
  \BibitemOpen
  \bibfield  {author} {\bibinfo {author} {\bibfnamefont {P.~M.~C.}\
  \bibnamefont {de~Oliveira}},\ }\href@noop {} {\bibfield  {journal} {\bibinfo
  {journal} {EPL (Europhysics Letters)}\ }\textbf {\bibinfo {volume} {119}},\
  \bibinfo {pages} {40007} (\bibinfo {year} {2017})}\BibitemShut {NoStop}%
\end{thebibliography}%
\end{document}